\documentclass{article}
\usepackage{textcomp} 

\usepackage{epsfig}
\usepackage{graphicx}
\usepackage[dvips]{color}
\usepackage{verbatim}
\usepackage{gensymb}
\usepackage{gensymb}
\usepackage{lineno}
\usepackage[margin=0.75in]{geometry}
\def\blfootnote{\xdef\@thefnmark{}\@footnotetext} 
\long\def\symbolfootnote[#1]#2{\begingroup%
\def\thefootnote{\fnsymbol{footnote}}\footnote[#1]{#2}\endgroup} 

\begin{document}
\setpagewiselinenumbers

%
\textbf{\large{Voluminous D$_2$ source for intense cold neutron beam production at the ESS.}} 

\vspace{1cm}
 E. Klinkby$^{1,2}$\symbolfootnote[1]{Corresponding author: esbe@dtu.dk}, K. Batkov$^{1}$, F. Mezei$^{1}$, T. {Sch\"onfeldt$^{1,2}$}, A. Takibayev$^{1}$, L.~Zanini$^{1}$   \vspace{0.5cm}

 \begin{flushleft}
1) European Spallation Source ESS AB, Box 176, S-221 00 Lund, Sweden\\
2) DTU Nutech, Technical University of Denmark, DTU Ris\o~Campus, Frederiksborgvej 399, DK-4000 Roskilde, Denmark\\
 \end{flushleft}
\vspace{1cm}

\section{Introduction}
The development of the flat moderator concept at ESS~\cite{batkov,mezei} recently opened up the possibility that a single flat moderator above the target could serve all the scattering instruments, that rely on high brightness. This would allow for the introduction of a fundamentally different moderator below the target for the complementary needs of certain fundamental physics experiments. To facilitate experiments depending on the total number of neutrons in a sizable beam, the option of a voluminous D$_2$ moderator, in a large cross-section extraction guide is discussed and its neutronic performance is assessed.


\section{Simulation setup}

\begin{figure}
\begin{minipage}{\linewidth}
\centering
\epsfig{figure=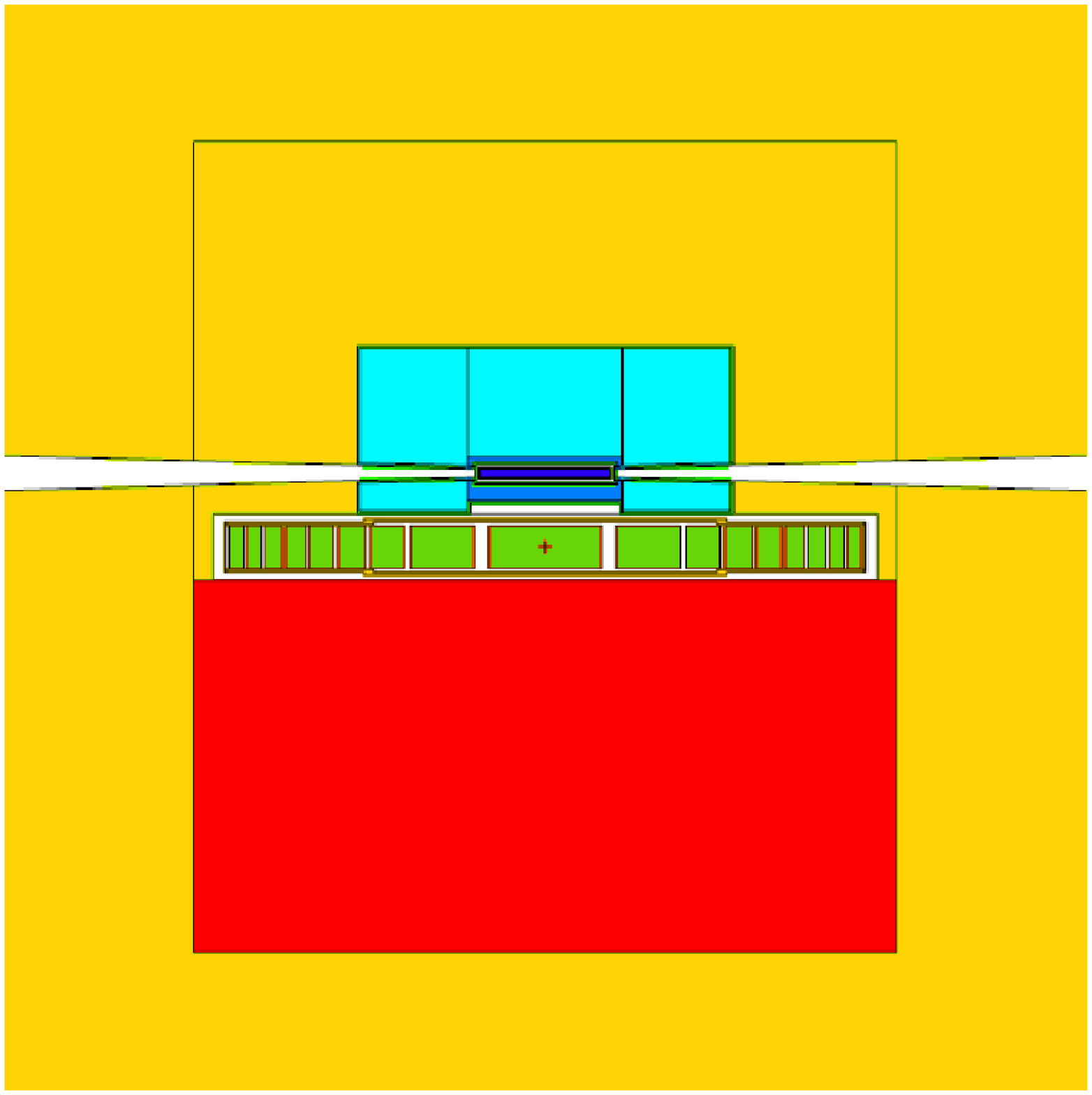,width=0.32\linewidth}
\epsfig{figure=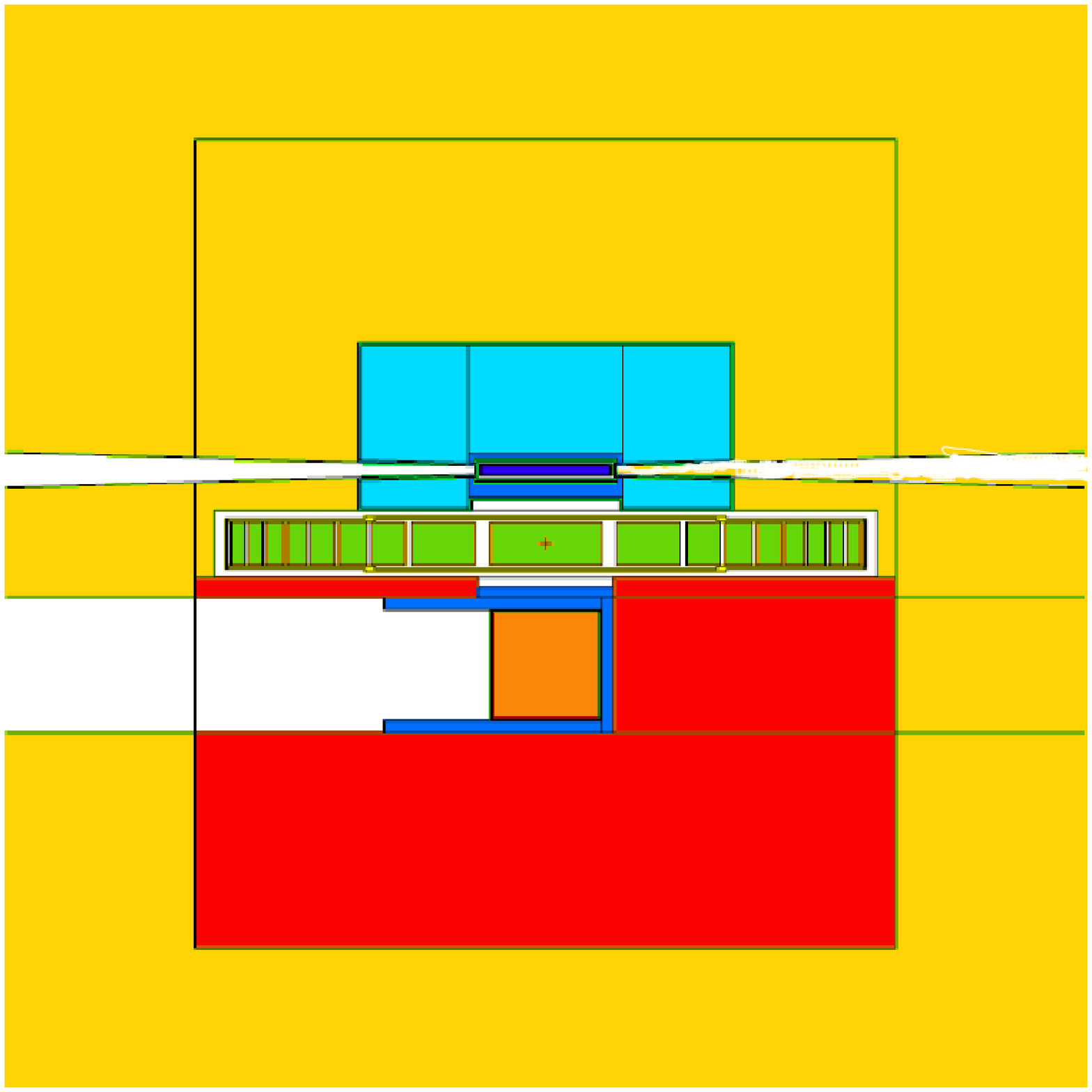,width=0.32\linewidth}
\epsfig{figure=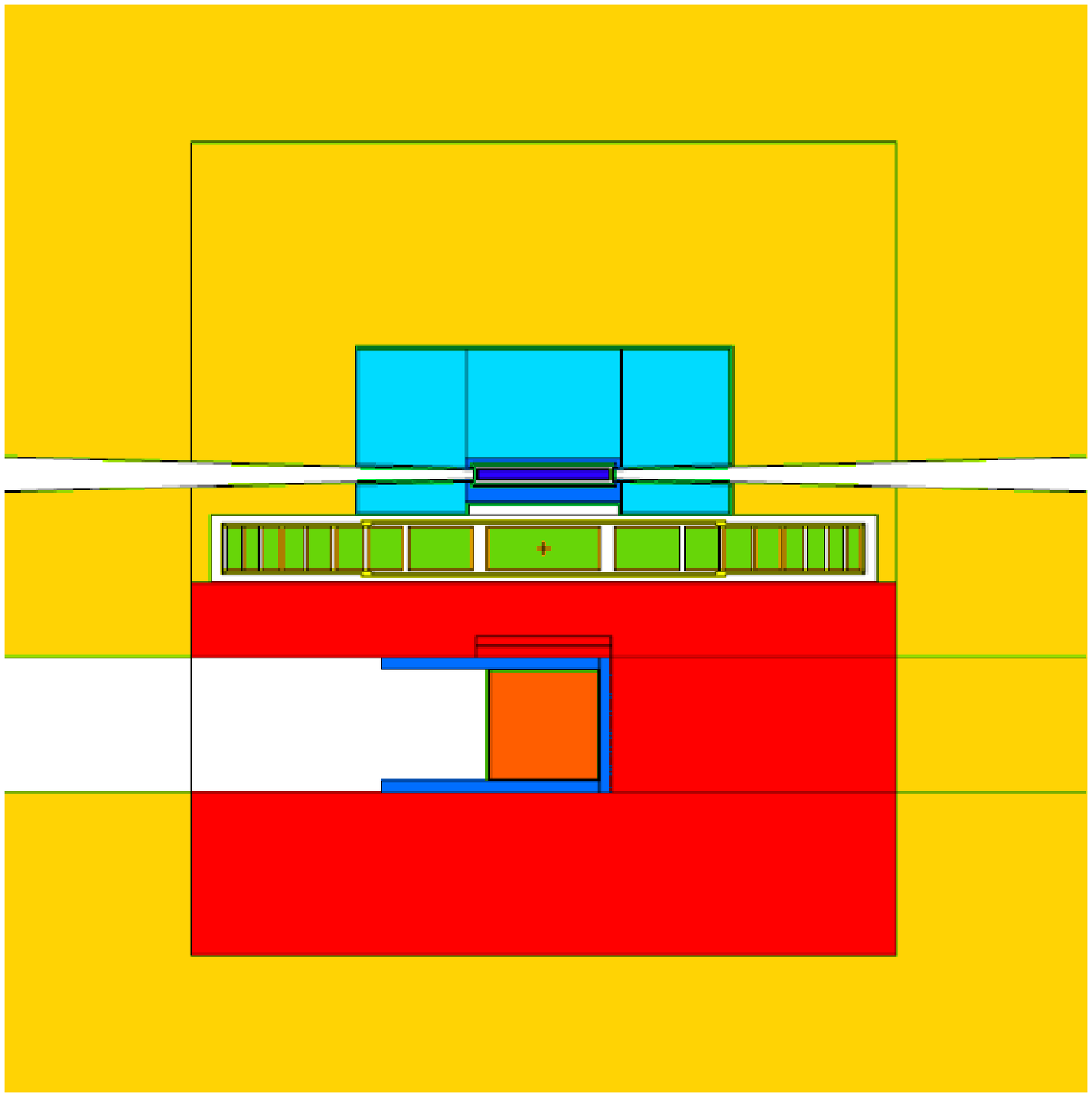,width=0.32\linewidth}
\put(-431,165){Reference}
\put(-440,140){Steel}
\put(-431,30){Pb}
\put(-421,101){Be}
\put(-267,165){Case 1a}
\put(-272,140){Steel}
\put(-267,30){Pb}
\put(-256,101){Be}
\put(-250,61){\tiny D$_2$}
\put(-100,165){Case 1b}
\put(-110,140){Steel}
\put(-97,30){Pb}
\put(-91,101){Be}
\put(-87,53){\tiny D$_2$}
\caption{Target, moderator and reflector geometry studied, corresponding to cases reference(left), 1a(middle) and 1b(right) discussed in the text.}
\label{fig:1a}
\end{minipage}\hfill
\end{figure}

The performance of a large liquid D$_2$ moderator (19~K, $\frac{1}{3}$ para-deuterium, $\frac{2}{3}$ ortho-deuterium) for high flux cold neutron production is studied under the assumption that a single (flat) para-hydrogen moderator above the target wheel serves the neutron scattering instruments relying on high brightness. Under this assumption, the region below the target wheel is, per default, completely filled with reflector and thus allow for a significant freedom for designing a voluminous D$_2$ moderator, with little or no consequences to the performance of the scattering instruments.\\
Large D$_2$ moderators are common at reactors and it was also selected as optimal for large intensity at the continuous spallation source SINQ. On short pulsed spallation sources the several millisecond long response time of these moderators make them disadvantageous for neutron scattering work.\\
As an example for demonstrating the performance feasibility, a $25$\thinspace cm$\times25$\thinspace cm$\times20.6$\thinspace cm rectangular D$_2$ moderator is placed centrally under the spallation hot-spot in a through-going beam tube with a $25$\thinspace cm$\times25$\thinspace cm cross-section. One end of the beam tube is left completely open to allow for neutron extraction whereas the other is foreseen to provide access to the D$_2$ moderator and would in addition contain the cooling piping and other service equipment. For simplicity, the simulations described below, assume that the access is plugged with the same material as the surrounding reflector, which is lead in the example studied here. The D$_2$ moderator is encapsulated in a 3mm aluminium casing and is on all sides except the viewed surface covered by a layer of water (for pre-moderation) of varying thickness (1-4~cm, most thick on the side facing the target). The simulation geometry is implemented in a MCNPX~\cite{Waters:2007zza,mcnp} model, as visualized in figure~\ref{fig:1a}.


To access the performance of the deuterium moderator, while monitoring the impact on the flat para-hydrogen moderator above, the following different cases are considered:
\begin{itemize}
\item {\bf Reference} : No D$_2$ moderator or extraction tube below the target. \\Figure~\ref{fig:1a}(left)
\item {\bf Case 1a}: D$_2$ moderator and extraction tube in up-most position, close to the target. Figure~\ref{fig:1a}(middle)
\item {\bf Case 1b}: D$_2$ moderator and extraction tube lowered by 10~cm. \\Figure~\ref{fig:1a}(right)
\end{itemize}


The setup described above is based on a lead reflector of 130~cm diameter, which has the advantage over beryllium (indispensable for optimal performance high brightness moderators) that it is less expensive for construction, operation and disposal. In addition, as fast neutron reflector below the target, lead has the potential to enhance the brightness of the moderator(s) above the target {(current estimate $\sim$ 10\%)}.

\section{Results}
\begin{figure}
\begin{minipage}{\linewidth}
\centering
\epsfig{figure=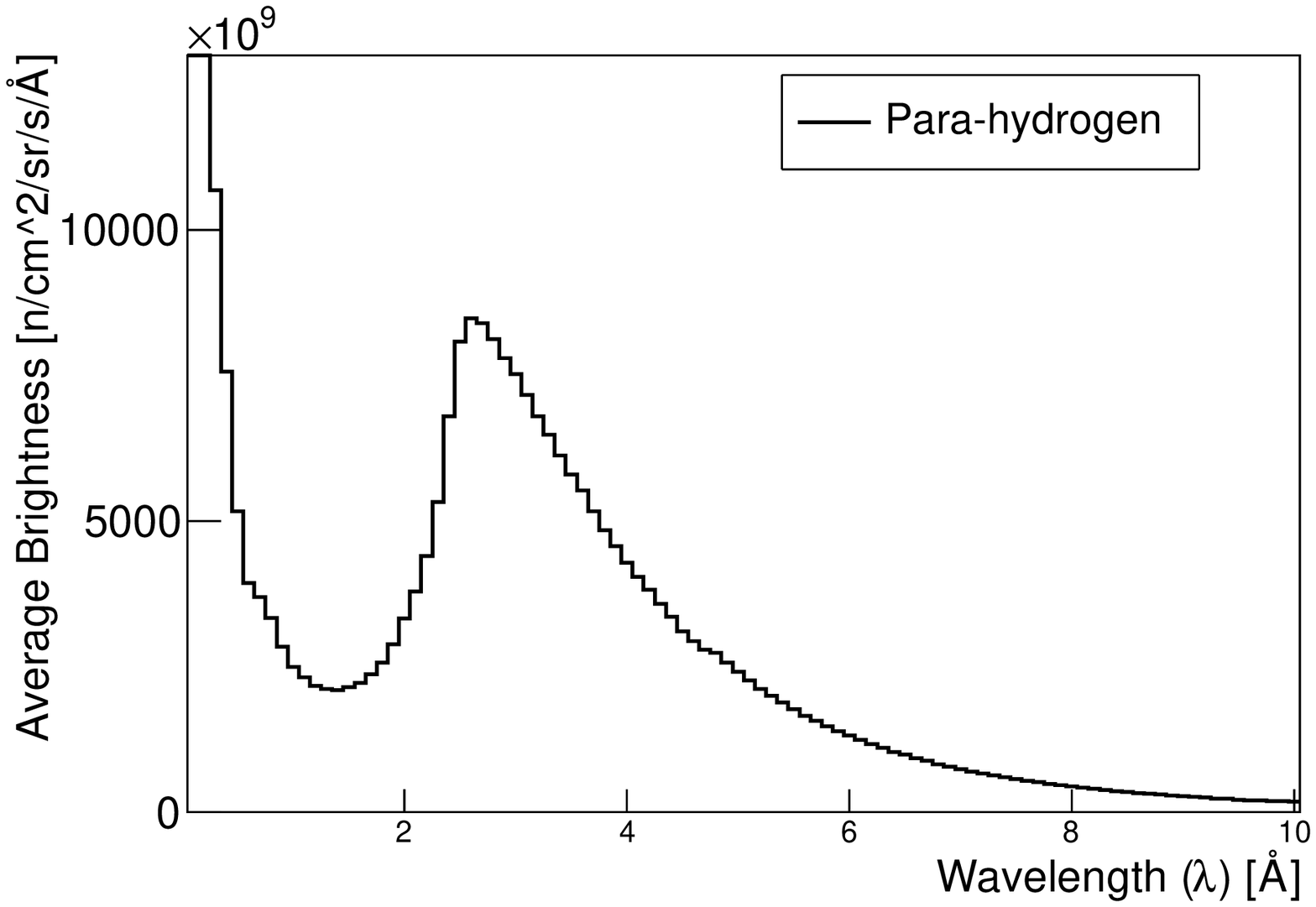,width=0.46\linewidth}
\epsfig{figure=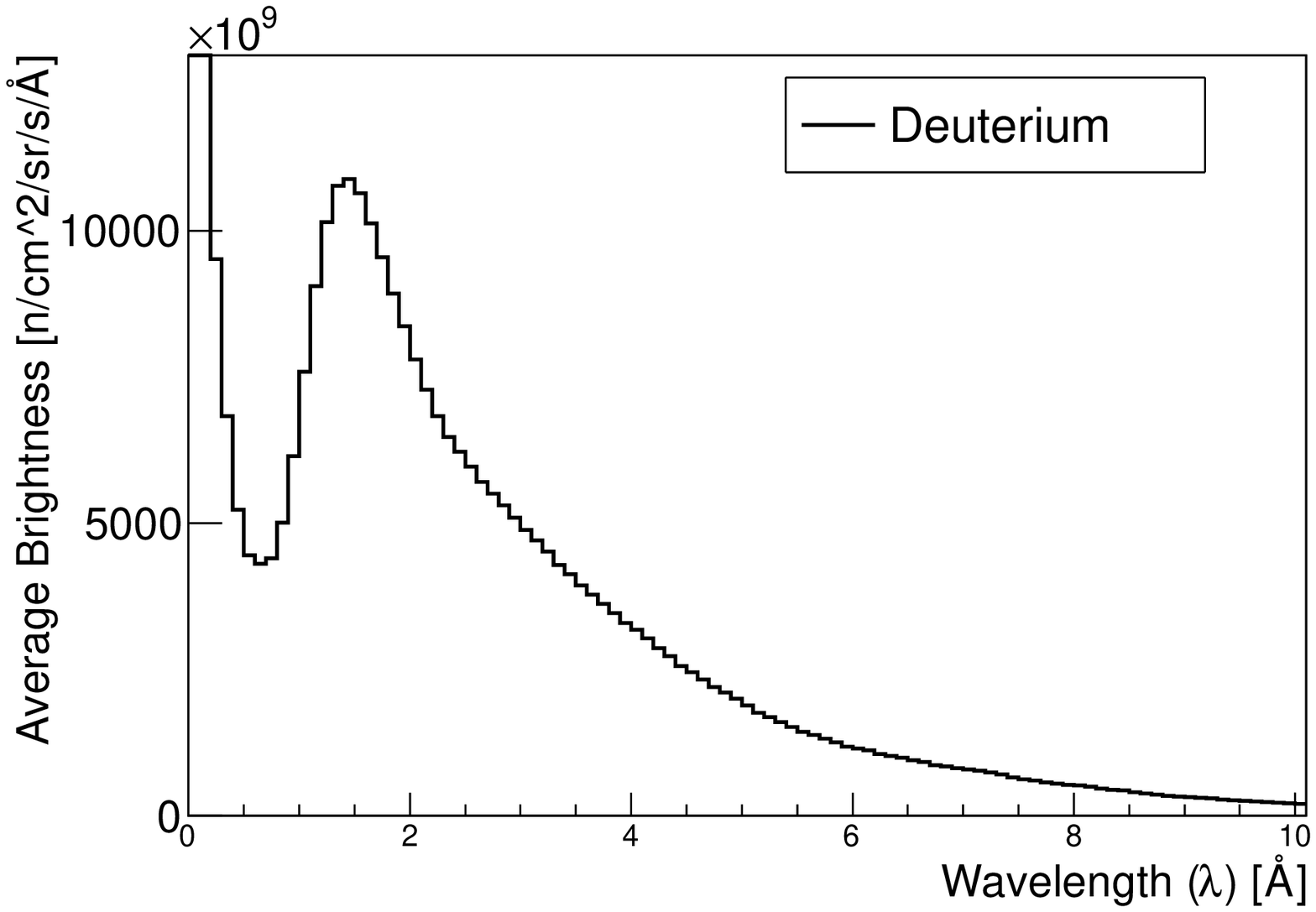,width=0.46\linewidth}
\caption{Left: Spectrum of the volume H$_2$ moderator of the TDR design~\cite{tdr}. Right: deuterium spectrum for Case 1a. In parts the brightness increase observed for deuterium with respect to the TDR para-hydrogen must be assigned to the introduction of lead as outer reflector as opposed to the steel based TDR outer reflector.}
\label{fig:specH}
\end{minipage}\hfill
\end{figure}

Figure~\ref{fig:specH}(left) shows brightness as a function of wavelength for the ESS baseline voluminous para-hydrogen moderator, whereas figure~\ref{fig:specH}(right) shows the corresponding curve of the D$_2$ moderator considered here (Case 1a)\footnote{Measured by point detectors (F5 tallies) placed at a distance of 5~m from the moderator surfaces, collimated so that only the moderator surface is viewed}. 
In each of the cases (reference, 1a,1b) the integrated cold brightness (0-5~meV) is given in table~\ref{tab:res}. This table shows that the scattering instruments viewing the para-hydrogen moderator, would suffer a $\sim$15\% brightness loss in case the deuterium moderator is at its up-most position (1a), compared to the reference, where the lead reflector below the target would contain no moderator and through-going beam tube. If the deuterium moderator is lowered by 10~cm (1b) this loss practically vanishes.

\begin{table}
  \begin{center}
  {\let\my=\\
    \begin{tabular}{l|c|c}
    Case& \multicolumn{2}{ |c }{Brightness [n/cm$^2$/sr/s]} \\
    \hline\hline
&Volume D$_2$ moderator (below)&Flat H$_2$ moderator (above) \\\hline
{\bf Reference}  & &3.34$\times 10^{13}$   \\\hline
{\bf 1a} & 6.83$\times 10^{12}$ & 2.80$\times 10^{13}$  \\\hline
{\bf 1b} & 4.56$\times 10^{12}$ & 3.22$\times 10^{13}$\\ 
    \end{tabular}
    \caption{Integrated: 0-5~meV cold neutron brightness from the upper, flat para-hydrogen (H$_2$) moderator of 1.5~cm height and the lower deuterium (D$_2$) moderator in the various studied cases. The brightness is evaluated from the flux calculated at a distance of 5~m from the viewed moderator surface and hence they represent the average over the viewed surface. The relative statistical uncertainties are $\sim$0.1\%.}
    \label{tab:res}
  }
  \end{center}
\end{table}

\section{Discussion}




For a number of experiments the key parameter determining the experimental reach is actually not the source brightness, $B$, but rather the total number of neutrons in the beam. This quantity is proportional to $A\times B$, where $A$ is the area of the viewed moderator surface and $B$ is the average brightness. Thus, it is useful to compare the performance of the $D_2$ moderator studied here, with the baseline H$_2$ moderator in terms of the quantity $A\times B$.
The results are shown in table~\ref{tab:discussion}. In this table, the corresponding result for the para-hydrogen volume moderator of the Technical Design Report~\cite{tdr} (TDR) is shown, assuming a $12$\thinspace cm$\times12$\thinspace cm beam extraction cross-section. 

\begin{table}
  \begin{center}
  {\let\my=\\
    \begin{tabular}{c|r r}
Case&$A\times B$ [n/sr/s] \\ \hline\hline
TDR H$_2$  - {\small $12~cm\times 12~cm$ } &1.17$\times 10^{15}$  \\ \hline
1a  D$_2$  - {\small $25~cm\times 20.6~cm$}& 4.27$\times 10^{15}$  \\ \hline
1b  D$_2$  - {\small $25~cm\times 20.6~cm$}& 2.85$\times 10^{15}$  \\ \hline
    \end{tabular}
    \caption{Neutron guide extraction cross-section multiplied by the integrated cold (0-5~meV) brightness from the deuterium (D$_2$) moderator in the various studied cases. For comparison, the same parameter is shown for the ESS baseline case (TDR - Technical Design Report~\cite{tdr}). The relative statistical uncertainties are all $\sim$0.1\%.}
    \label{tab:discussion}
  }
  \end{center}
\end{table}

Table~\ref{tab:discussion} shows that for fundamental neutron beam experiments (as opposed to neutron scattering work), the larger D$_2$ moderator studied here as a first trial considerably outperforms the smaller TDR moderator in both cases (1a) and (1b). The optimization of this large cross-section moderator option for delivering highest number of neutrons for neutron beam experiments is in progress. The optimization concerns parameters such as lay-out dimensions, moderator material, pre-moderator geometry, choice of reflector or reflector combination. It can be expected that the 3-4 fold gain potential demonstrated here (table~\ref{tab:discussion}) over the highest beam intensity achievable by moderators designed for neutron scattering work can be further enhanced significantly.



\section{Conclusions}
Under the assumption, that the scattering instruments at the ESS are served by a single moderator above the target, the performance of a $25$~cm$\times25$~cm$\times20.6$~cm rectangular D$_2$ moderator placed in a through-going beam-tube under the target has been studied. The results prove the feasibility of providing by this approach at least 3-4 times enhanced cold neutron beam intensity for fundamental physics experiments, such as $n\bar{n}$ oscillations without essentially impacting the performance of the moderators serving the neutron scattering beam lines. It is expected that the performance, as well as the impact on the performance of the upper para-hydrogen moderator would benefit significantly from the ongoing optimization work, including the enhancement of brightness of the upper moderators by judicious choice of the reflector material $/$ configuration below the target.

\bibliographystyle{elsarticle-num}
\bibliography{mybib}


\end{document}